\documentclass[twoside,11pt]{article}

\usepackage{obs_study_style}
\usepackage{amsmath}
\usepackage{booktabs}
\usepackage{listings}
\usepackage{color}
\usepackage[skip=0pt plus0pt, indent=16pt]{parskip}
\definecolor{codegreen}{rgb}{0,0.65,0}
\definecolor{codegray}{rgb}{0.5,0.5,0.5}
\definecolor{codepurple}{rgb}{0.58,0,0.82}
\definecolor{backcolour}{rgb}{0.95,0.95,0.95}
\lstdefinestyle{mystyle}{
    backgroundcolor=\color{backcolour},   
    commentstyle=\color{codegreen},
    keywordstyle=\color{black},
    numberstyle=\tiny\color{white},
    stringstyle=\color{blue},
    basicstyle=\ttfamily\footnotesize,
    breakatwhitespace=false,         
    breaklines=true,                 
    captionpos=b,                    
    keepspaces=true,                 
    numbers=left,                    
    numbersep=5pt,                  
    showspaces=false,                
    showstringspaces=false,
    showtabs=false,                  
    tabsize=2
}
\newcommand{\bX}{\boldsymbol{X}}
\newcommand{\btheta}{\boldsymbol{\theta}}
\newcommand{\bF}{\mathbb{F}}

\newcommand{\Pro}{\text{P}}
\newcommand{\roc}{\text{ROC}}
\newcommand{\roch}{\widehat{\text{ROC}}}

\newcommand{\E}{\text{E}}
\newcommand{\bbeta}{\boldsymbol{\beta}}
\newcommand{\ecd}{\text{ECD}_2}

\heading{}{}{}{}{}{Nathaniel P. Dowd, James Chappell, Natasha B. Halasa, and Andrew J. Spieker}

\ShortHeadings{Receiver operating characteristic methods}{Dowd et al.}
\firstpageno{1}

\begin{document}

\title{An overview of methods for receiver operating characteristic analysis, with an application to SARS-CoV-2 vaccine-induced humoral responses in solid organ transplant recipients}

\author{\name Nathaniel P. Dowd \email nathaniel.p.dowd@vanderbilt.edu \\
       \addr Department of Biostatistics\\
       Vanderbilt University\\
       Nashville, TN 37203, USA
       \AND
       \name Bryan Blette \email bryan.blette@vumc.org \\
       \addr Department of Biostatitics\\
       Vanderbilt University Medical Center\\
       Nashville, TN 37203, USA
       \AND
       \name James D. Chappell \email jim.chappell@vumc.org\\
       \addr Division of Pediatric Infectious Diseases\\
       Vanderbilt University Medical Center\\
       Nashville, TN 37232, USA
       \AND
       \name Natasha B. Halasa \email natasha.halasa@vumc.org\\
       \addr Division of Pediatric Infectious Diseases\\
       Vanderbilt University Medical Center\\
       Nashville, TN 37232, USA
       \AND
       \name Andrew J. Spieker \email andrew.spieker@vumc.org \\
       \addr Department of Biostatitics\\
       Vanderbilt University Medical Center\\
       Nashville, TN 37203, USA
       }

\maketitle

\begin{abstract}

Receiver operating characteristic (ROC) analysis is a tool to evaluate the capacity of a numeric measure to distinguish between groups, often employed in the evaluation of diagnostic tests. Overall classification ability is sometimes crudely summarized by a single numeric measure such as the area under the empirical ROC curve. However, it may also be of interest to estimate the full ROC curve while leveraging assumptions regarding the nature of the data (parametric) or about the ROC curve directly (semiparametric). Although there has been recent interest in methods to conduct comparisons by way of stochastic ordering, nuances surrounding ROC geometry and estimation are not widely known in the broader scientific and statistical community. The overarching goals of this manuscript are to (1) provide an overview of existing frameworks for ROC curve estimation with examples, (2) offer intuition for and considerations regarding methodological trade-offs, and (3) supply sample R code to guide implementation. We utilize simulations to demonstrate the bias-variance trade-off across various methods. As an illustrative example, we analyze data from a recent cohort study in order to compare responses to SARS-CoV-2 vaccination between solid organ transplant recipients and healthy controls.

\end{abstract}


\section{Introduction}

Receiver operating characteristic (ROC) analysis is a tool to summarize the extent to which a numeric measure accurately distinguishes between groups. ROC methodology has historical roots in electrical engineering as a method to detect enemy objects in battlefields, but became heavily associated with assessment of diagnostic utility shortly after its development (\citealp{Green66}). In clinical practice, it is often of interest to determine whether a diagnostic test based on some numeric measure has the ability to discriminate between two states of health. In particular, it is often a goal to develop classification measures that help reduce the invasiveness and/or costs associated with diagnostic tests while maintaining accuracy of classification. For example, a ``gold standard'' test in the diagnosis of prostate cancer as determined by biopsy is invasive and impractical as a routine clinical screening measure; the prostate-specific antigen is a biomarker that has been used for prostate cancer detection (\citealp{Barry01}). Numerous examples of ROC analysis have been published, and its use is still widespread in the modern era (\citealp{Leblhuber88, Spieker13, Johnson16, Xu24, Zhu24, Zhou24}). Summary measures associated with ROC analysis have also long been used to compare classification ability across machine learning methods (\citealp{Hanley83}).

More recently, researchers have more broadly recognized stochastic ordering as a scientifically relevant measure in its own right to compare groups---including in contexts outside of diagnostic tests (\citealp{Acion06, Zhang19, Spieker21}). Proper procedures for model selection, implementation, inference, and interpretation are less widely known in the statistical and scientific community as compared to, for example, methods to compare means across population strata (\citealp{Freedman09}).

Parametric, nonparametric, and semiparametric approaches to ROC estimation have been characterized in the literature, largely separately (\citealp{Metz86, Thompson89, Metz98, Pepe00, Pepe03}). A bias-variance trade-off is to be expected when comparing estimators across these classes (with more structure generally offering greater efficiency under correct assumptions, but introducing bias under assumption violations), though the nature of that trade-off in the space of ROC methods is not as well explored. Further, inference regarding the ROC curve is deceptively simple due to the fact that both the \textit{x}- and \textit{y}-coordinates of the overall curve are estimated. Subtleties and nuances regarding the fundamental geometry of ROC-based methods, as well as considerations surrounding estimation, merit more explicit characterization and further illustration. The fundamental goal of this manuscript is to address this need.

Throughout this manuscript, we anchor our illustrations using two foundational examples for ease of characterizing various trade-offs: exponential methods and normal methods. The remainder of this manuscript is organized as follows. In Section 2, we provide definitions and an overview of the ROC curve and considerations regarding its geometry. In Section 3, we discuss and provide sample R code for estimation procedures associated with the ROC curve, including related summary measures. In Section 4, we describe and present the results of various simulation studies to highlight the trade-offs between approaches. In Section 5, we apply several methods to a recent study seeking to characterize immune responses to SARS-CoV-2 vaccination in solid organ transplant (SOT) recipients. Finally, we conclude with a discussion of our findings in Section 6.

\section{The ROC curve and its geometry}

\subsection{Notation}

Throughout this manuscript, we let $i = 1, \dots, n_0$ index independently sampled observations from a reference distribution with outcomes $Y_{01}, \dots, Y_{0n_0}$, each having common cumulative distribution function (CDF) $F_0(t) = \Pro(Y_0 \leq t)$, survivor function $S_0(t) = 1 - F_0(t)$, and quantile function $F_0^{-1}(p)$. Analogously, we let $j = 1, \dots n_1$ index independently sampled observations from a comparator distribution with outcomes $Y_{11}, \dots, Y_{1n_1}$, each having common CDF $F_1(t)$, survivor function $S_1(t)$, and quantile function $F_1^{-1}(p)$. Note that in cases where $F_0$ and $F_1$ are not invertible (e.g., associated with discrete-valued outcomes), $F_0^{-1}(\cdot)$ and $F_1^{-1}(\cdot)$ denote generalized inverses (for instance, $F_0^{-1}(p) = \inf\lbrace t : F_0(t) > p\rbrace$). For ease of notation, we let $n = n_0 + n_1$ denote the combined sample size, and we will often refer generically to the measure of interest as $Y$.

The utility of the measure $Y$ as a classifier is most easily described first at a particular cut-off point, $c \in \text{supp}(Y)$. The true positive rate (TPR) marks the proportion of observations in the comparator group correctly classified at that cut-off; the false positive rate (FPR) marks the proportion of observations in the reference group incorrectly classified. The graph of an ROC curve summarizes these two complementary classifying features of the measure $Y$ across all possible cut-off points, and is defined as a plot of the FPR (i.e., one minus the specificity) at a particular cut-off point on the \textit{x}-axis and the TPR (i.e., sensitivity) at that cut-off on the \textit{y}-axis. The ROC curve is a monotone, surjective function $\roc : [0, 1] \longrightarrow [0, 1]$, so that $p$, the FPR, denotes the input and $\roc(p)$, the TPR, denotes the output (\citealp{Pepe00, Fawcett06}).

\subsection{Conventions and symmetries}

To characterize the ROC curve as a function of observable data, we must specify whether higher or lower values of $Y$ signify a less desirable outcome. In many settings, scientific knowledge will provide a researcher with this information \textit{a priori}: for example, higher hemoglobin A1c (HbA1c) is generally known to be less desirable in patients with type 2 diabetes mellitus (\citealp{Cavero17}), and lower CD4$^{+}$ T cell count is known to be less desirable in patients with human immunodeficiency virus (\citealp{Doitsh14}). When lower values of $Y$ are considered less desirable, the ROC function is best defined as $\roc(p) = F_1(F_0^{-1}(p))$, for $0 \leq p \leq 1$. On the other hand, when higher values of $Y$ are considered less desirable, the ROC function is instead best defined as $\roc(p) = S_1(S_0^{-1}(p))$. In either case, the ROC curve can be thought of as a sensible composition of functions: the inner function determines the cut-off value, $c_p$, such that $100 \times (1 - p)\%$ of those in the reference group are correctly classified, and the outer function determines the relative frequency of those in the comparator group correctly classified at said cut-off. This is illustrated in Figure 1.

\begin{figure}[h!]
    \centering
    \includegraphics[width=1\linewidth]{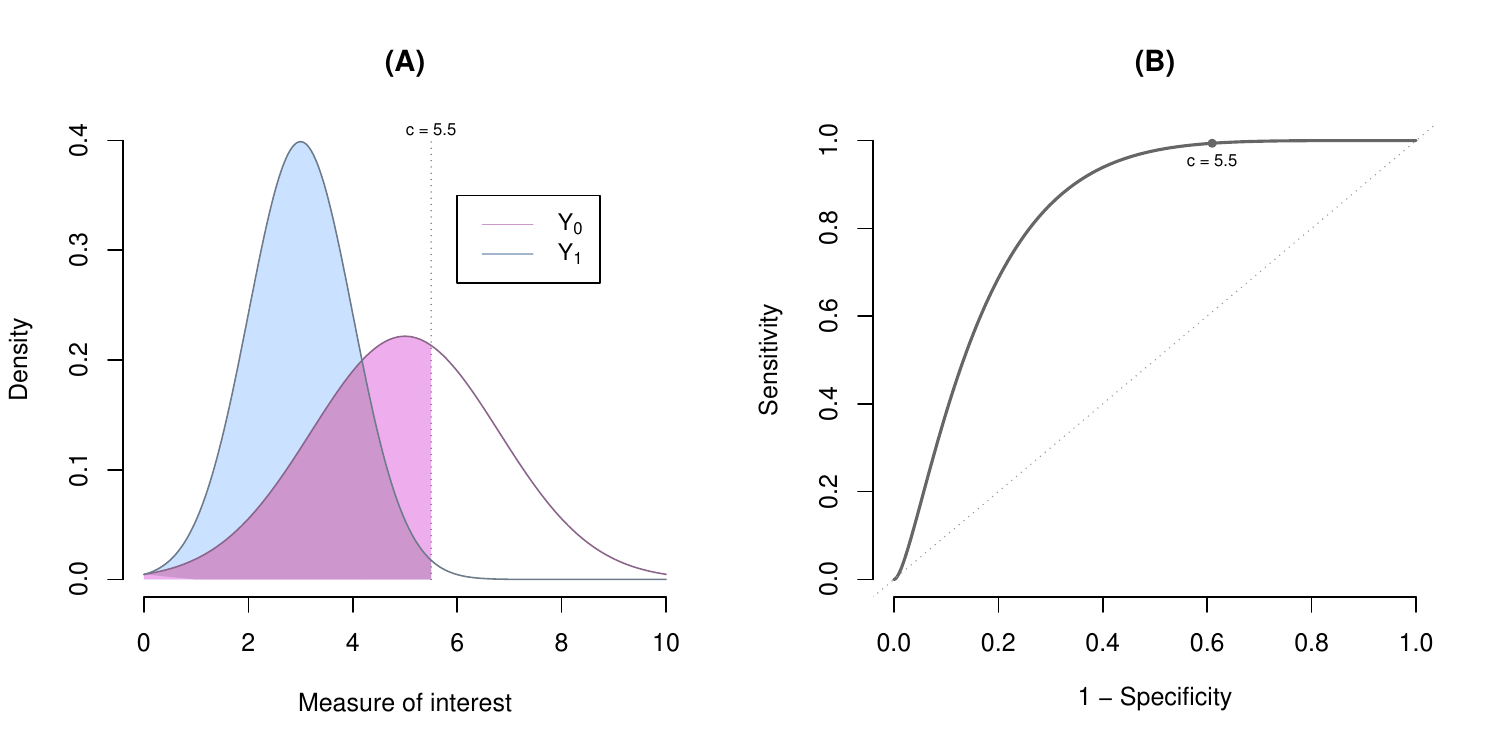}
    \caption{Illustration of the ROC curve under the convention that lower values are less desirable. In Panel (A), the reference group has a higher mean value of $Y$ and substantially greater variability; a cut-off point of $c=5.5$ results in a FPR of approximately 61\% (shaded pink) and a TPR of approximately 99\% (shaded blue). Panel (B) depicts the full ROC curve, which can be thought of as sliding the cut-off point across all possible values of $c$ (the specific point corresponding to $c=5.5$ is marked on the ROC curve). Increasing values of $c$ correspond to sliding ``up and to the right'' on the ROC curve, while decreasing values of $c$ correspond to sliding ``down and to the left.''}
    \label{fig:Fig1}
\end{figure}

In addition to setting a directional convention, we must also decide which group serves as the reference group (0) and which is the comparator (1). This will often have a natural synergy with the convention of direction described above (for example, healthy controls will tend to have lower HbA1c as compared to patients with type 2 diabetes). However, in a comparative study of two active treatments (e.g., the effect of proton vs. photon therapy for cancer treatment), context will not always allow the researcher to so easily intuit which group should serve as a reference in advance (\citealp{Baumann20}). Further, it may not even always be possible to characterize a higher value of $Y$ as ``better'' or ``worse'' (e.g., a comparison of hormone levels between males and non-males), such that two independent researchers analyzing the same data could reasonably choose opposite conventions (\citealp{Thakur09}). To maintain the typical convention of an ROC curve that, loosely speaking---and with further elaboration in Section 2.3---lies mostly on the upper-left side of the unit square $[0, 1] \times [0, 1]$, a change in convention of directionality necessitates a simultaneous reversal of groups. This translates to an important symmetry in the resulting ROC curve, described as follows: a change in directional convention corresponds to the mapping $(x, y) \mapsto (1 - x, 1 - y)$, and a reversal of groups corresponds to the mapping $(x, y) \mapsto (y, x)$, so that if $(p, \roc(p))$ denotes the graph of the original ROC curve, the graph of the updated ROC curve is denoted by $(1 - \roc(p), 1 - p)$. This phenomenon is illustrated in Figure 2.

\begin{figure}[h!]
    \centering
    \includegraphics[width=0.7\linewidth]{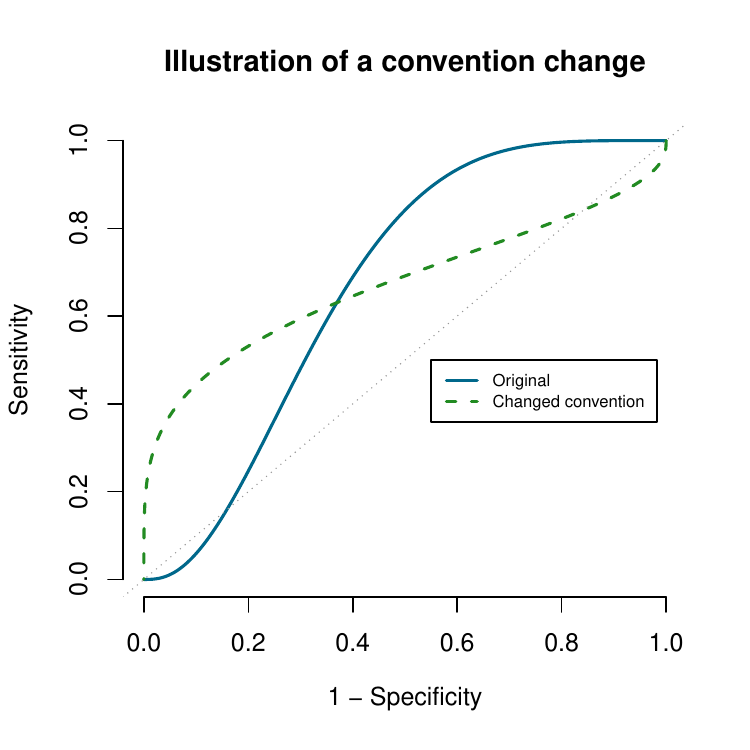}
    \caption{Illustration of an ROC convention change. An ROC curve (shown in solid blue) is depicted so that it, loosely speaking, ``lies mostly on the upper-left side.'' Importantly, consideration of stochastic ordering can be of scientific interest even when there is no natural way to define ``better'' or ``worse.'' Following the symmetries associated with a change in directionality and a reversal of groups, the ROC curve under a change in convention (dashed green) can be realized as a reflection of the original curve about the line $y = 1 - x$.}
    \label{fig:Fig2}
\end{figure}

In line with our ultimate motivating example (Section 5), we will follow the convention that higher values are more desirable (largely without loss of generality, with a notable exception serving as the topic of discussion in Section 3.4).

\subsection{Area under the curve}

The overall quality of $Y$ as a classifier can be summarized by the area under the ROC curve (AUC), given by:
\begin{eqnarray*}
\text{AUC} = \int_0^1 \roc(p) dp = \int_0^1 F_1(F_0^{-1}(p)) dp = \Pro(Y_0 > Y_1).
\end{eqnarray*}
The value of the AUC is bounded below by 0 and above by 1; a change of directional convention alone or a reversal of groups alone would each update the AUC via the relationship $\text{AUC}^{\prime} = 1 - \text{AUC}$, which is to say that the AUC is preserved by the convention change illustrated in Figure 2; typically, a convention is chosen such that the AUC is no smaller than $0.5$. An ROC curve for two well-separated distributions will pass through the point (0, 1), with an AUC of 1.0 (\citealp{Fawcett06}). An ROC curve for two identical distributions will lie along the line $y=x$, with an AUC of 0.5. Importantly, equality of $F_0(t)$ and $F_1(t)$ for all values of $t$ implies an AUC of 0.5, but not the other way around (i.e., it is possible to achieve an AUC of 0.5 when comparing two distributions that are different; an example will be provided in Section 2.5). This distinction is important, as there are two fundamental tests that may be of interest when evaluating the predictive capacity of $Y$: a test of a \textit{weak} null hypothesis, $H_0 : \text{AUC} = 0.5$, and a test of a \textit{strong} null hypothesis, $H_0 : \roc(p) = p$ for all $0 \leq p \leq 1$ (i.e., the ROC curve lies along the line $y=x$). Of note, and as previously pointed out (\citealp{Hart01}), a test of the weak null hypothesis is sometimes mischaracterized as a test of a difference in medians: it can be shown that two groups can have the same median value of $Y$ when the weak null is false, and that two groups can have different median values of $Y$ when the weak null is true. Although not our emphasis, we offer methodological guidance on both characterizing and testing the weak and the strong null hypotheses.

\subsection{ROC curves emerging from exponential data}

Suppose that $Y_0 \sim \text{Exponential}(\lambda_0)$, with CDF given by $F_0(t) = 1 - \exp(-\lambda_0 t)$ and quantile function given by $F_0^{-1}(p) = -\lambda_0^{-1}\log(1 - p)$; analogously, assume that $Y_1 \sim \text{Exponential}(\lambda_1)$. It is readily shown in this setting that the family of ROC curves indexed by this parameterization is given by $\roc(p) = 1 - (1 - p)^{\alpha}$, where $\alpha = \lambda_1/\lambda_0 \geq 1$. We choose to refer to this family of ROC curves as the \textit{biexponential} ROC family in order to follow the previously established terminology used for normal data (Section 2.5). This family is illustrated in Figure 3(A) for different choices of $\alpha$. It is straightforward to show that the AUC in this example can be expressed in closed form as $\text{AUC}=1-(\alpha+1)^{-1}$, $\alpha \geq 1$, and that both the weak and strong null hypotheses are represented by $H_0 : \alpha = 1$. A convention change as described in Section 2.2 and illustrated in Figure 2 updates the form of the ROC curve as $\roc(p) = p^{\alpha^{\prime}}$, $0 < \alpha^{\prime} < 1$, which, quite notably, indexes a different family of ROC curves than the family indexed by the original convention. Together, these facts signify a certain restrictiveness in the form of the single-parameter biexponential ROC family.

\subsection{ROC curves emerging from normal data}

Suppose now that $Y_0 \sim \mathcal{N}(\mu_0, \sigma_0^2)$, with CDF given by $F_0(t) = \Phi((t - \mu_0)/\sigma_0)$ and quantile function given by $F_0^{-1}(p) = \mu_0 + \sigma_0\Phi^{-1}(p)$; analogously, assume that $Y_1 \sim \mathcal{N}(\mu_1, \sigma_1^2)$. It is straightforward to show that the family of ROC curves indexed by this parameterization is given by $\roc(p) = \Phi(\delta\sigma_1^{-1} + \sigma_0\sigma_1^{-1}\Phi^{-1}(p))$, where $\delta = \mu_0 - \mu_1$. This family is often referred to as the \textit{binormal} ROC family (\citealp{Hanley88}), and is illustrated in Figures 3(B) and 3(C) for different choices of $\delta$ and $\sigma_0/\sigma_1$. For reasons that will become more apparent when we discuss estimation, the binormal family is often expressed as $\roc(p) = \Phi(\beta_0 + \beta_1\Phi^{-1}(p))$, with $\beta_0 = \delta\sigma_1^{-1}$ and $\beta_1 = \sigma_0\sigma_1^{-1}$. The binormal ROC curve offers certain flexibility not offered by the biexponential family. First, note that although no closed-form expression exists for the AUC of a binormal ROC curve, the weak null is represented by $H_0 : \beta_0 = 0$ (that is, $\mu_0 = \mu_1$), and the strong null is represented by $H_0 : \beta_0 = 0 \text{ and } \beta_1 = 1$ (that is, $\mu_0 = \mu_1$ and $\sigma_0 = \sigma_1$). Secondly, note that a convention change corresponds to a re-parameterization that remains in the same binormal family under the original convention, with $\beta_0 \mapsto \beta_0/\beta_1$ and $\beta_1 \mapsto -1/\beta_1$. The latter fact, in particular, speaks to the broader range of shapes that can be taken by the binormal family under a single convention.

\begin{figure}[h!]
    \centering
    \includegraphics[width=0.325\linewidth]{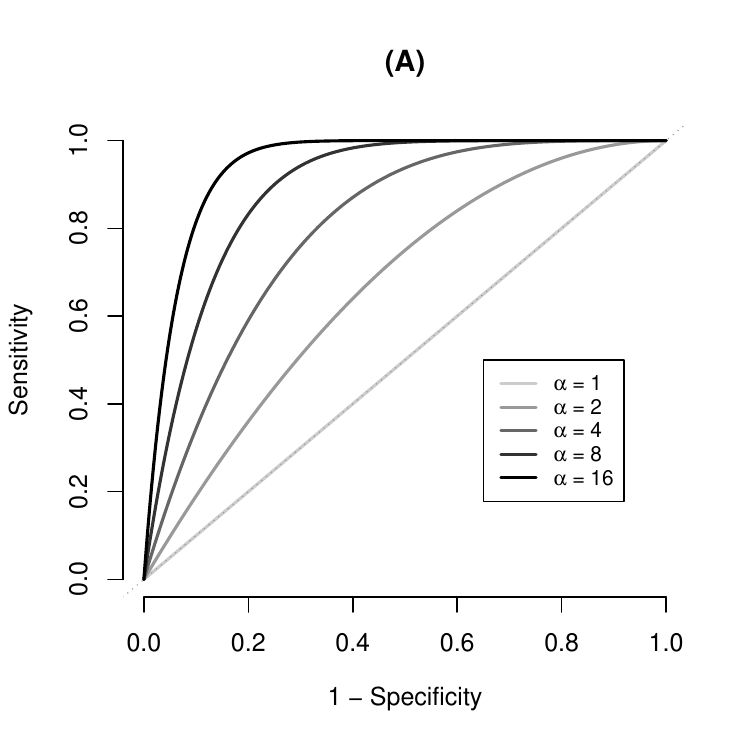}
    \includegraphics[width=0.325\linewidth]{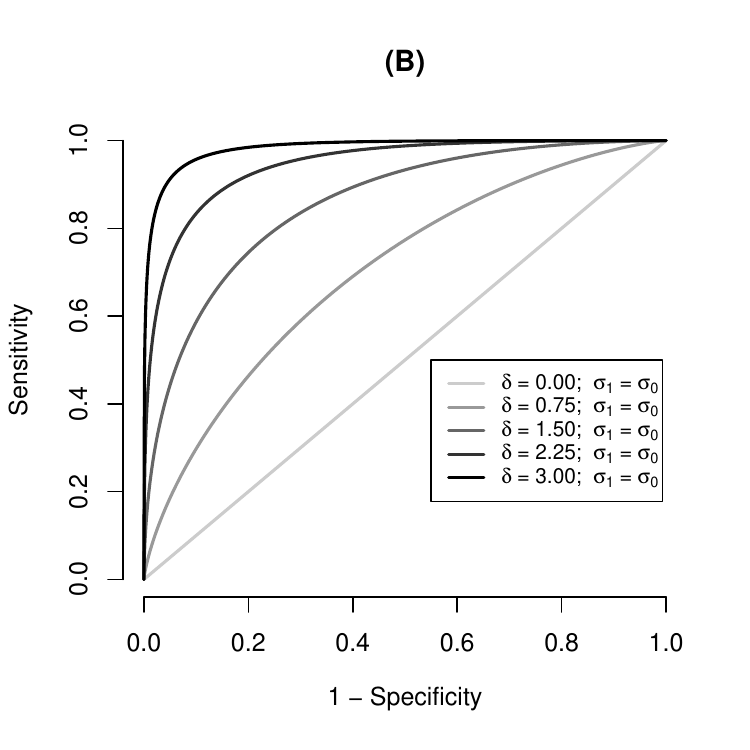}
    \includegraphics[width=0.325\linewidth]{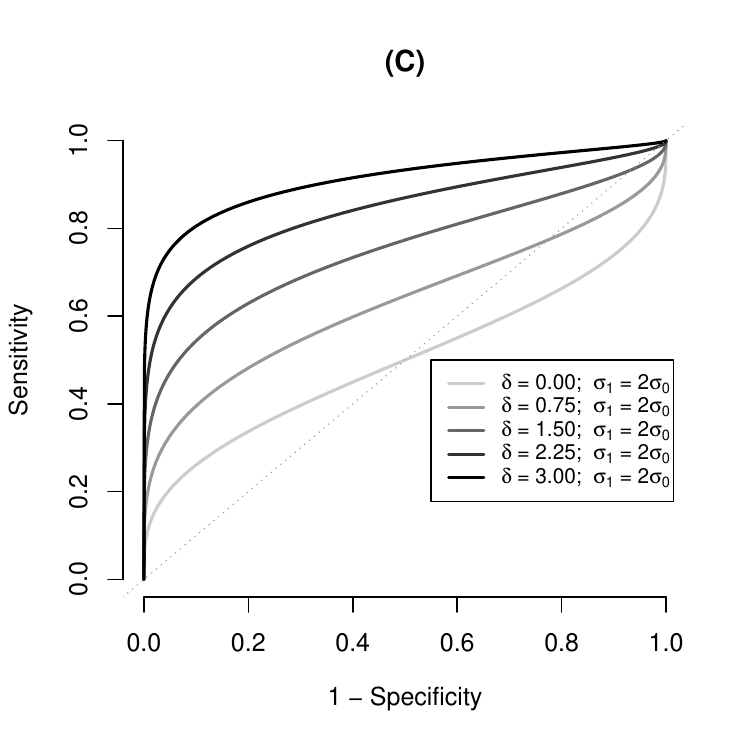}
    \caption{Illustration of families of ROC curves. Panel (A) illustrates a family of biexponential ROC curves; note that for $\alpha > 1$, the ROC curve always adheres to the same class of asymmetry about $y=-x$ under a chosen convention. Panel (B) illustrates a family of binormal ROC curves indexed by normal data with equal variances (i.e., $\beta_1 = 1$); Panel (C) illustrates a family of binormal ROC curves indexed by normal data with unequal variances (in particular, $\beta_1 = \sigma_0/\sigma_1 = 0.5$; note that choosing $\beta_1 = -2$ would result in a family of curves reflected about the line $y=-x$ for specific values of $\delta$; not depicted).}
    \label{fig:Fig3}
\end{figure}

\subsection{More on the biexponential and binormal curves}

We have now derived two forms for the ROC curve: one that emerges from exponentially distributed data (Section 2.4) and one that emerges from normally distributed data (Section 2.5). Importantly, ROC curves taking either of these forms can emerge from other combinations of distributions. That is to say, for example, that exponentially distributed data imply a biexponential form for the ROC curve, but not the reverse. Consider the following examples. Suppose first that $Y_0$ follows a normal distribution with quantile function $F_0^{-1}(p) = \mu_0 + \sigma_0\Phi^{-1}(p)$. It is straightforward to verify that a biexponential ROC curve is achieved by assigning the following CDF to $Y_1$: $F_1(t) = 1 - (1 - \Phi((t - \mu_0)/\sigma_0))^{\alpha}$. As another example, suppose that $Y_0$ follows an exponential distribution with quantile function $F_0^{-1}(p) = -\lambda_0^{-1}\log(1 - p)$. It is again straightforward to verify that a binormal ROC curve is achieved by assigning the following CDF to $Y_1$: $F_1(t) = \Phi(\beta_0 + \beta_1\Phi^{-1}(1 - \exp(-\lambda_0 t)))$. The CDFs that induce a desired structure on the form of the ROC curve do not go by known names, and are unlikely to offer practical relevance except to illustrate the lack of a one-to-one relationship between parametric assumptions on the data and the form of the ROC family. Nevertheless, the idea that multiple parametric structures on the data can produce the same family of ROC curves is crucial to understand why direct modeling of the ROC curve without parametric assumptions on the underlying data can be conceptualized as a semiparametric method. These examples offer a heuristic but nevertheless constructive argument that approaches to directly model the ROC curve do not implicitly impose a single parametric structure on the data (which, if they did, would render the direct modeling approaches trivial).

\section{Estimation of the ROC curve}

In this section, we outline various methods for estimation of the ROC curve. For ease of exposition, we assume for the majority of our examples that we are interested in obtaining point and interval estimates for the sensitivity, $\roc(p)$, at given values of the FPR, $p$. In Section 3.4, we briefly discuss alternative ways of characterizing uncertainty.

\subsection{Nonparametric estimation}

Let $\bF_0(t) = n_0^{-1}\sum_{i = 1}^{n_0} I(Y_{0i} \leq t)$ denote the empirical CDF for the reference group, and $\bF_0^{-1}(p) = \inf\lbrace t : \bF_0(t) > p \rbrace$; define $\bF_1(t)$ and $\bF_1^{-1}(p)$ analogously. The empirical ROC curve can be estimated nonparametrically as $\roch(p) = \bF_1(\bF_0^{-1}(p))$. In a finite sample, the empirical ROC curve takes the shape of a staircase. Given a set of data, the empirical ROC curve can be generated using the R code presented in Code Block 1.

\begin{lstlisting}[language=R]
## Code Block 1: Nonparametric Estimation of the ROC curve

## values : string of values for the measure of interest
## group  : string of indicators for group membership
compute_roc <- function(values, group, plot = "TRUE") {
    sorted_values <- values[order(values)]
    sorted_group <- group[order(values)]
    cumsum_pos <- cumsum(sorted_group == 1)
    cumsum_neg <- cumsum(sorted_group == 0)
    n_pos <- sum(group == 1)
    n_neg <- sum(group == 0)
    TPR <- cumsum_pos/n_pos
    FPR <- cumsum_neg/n_neg
    TPR <- c(0, TPR, 1)
    FPR <- c(0, FPR, 1)
    if (plot == "TRUE") {
    plot(FPR, TPR, type = "l", xlim = c(0,1), ylim = c(0,1),
         frame.plot = FALSE, lwd = 2, ylab = "Sensitivity",
         xlab = expression(paste("1 ", "\u2013", " Specificity")),
         main = "Empirical ROC Curve")
    }
    return(list(FPR = FPR, TPR = TPR))
}
## output : ordered pairs to be plotted on ROC curve
\end{lstlisting}

Constructing point-wise confidence bands is deceptively simple. We are oriented to think in terms of conditioning on the values marked on the \textit{x}-axis and characterizing uncertainty in corresponding values on the \textit{y}-axis (e.g., forming point-wise confidence bands in linear regression). However, the values marked on both the \textit{x}- and \textit{y}-axes are the result of estimation at fixed values of a cut-off point. The temptation may be, since the point-wise confidence bands involve vertical slices, to form a confidence interval for a one-sample proportion (e.g., a Wilson interval) for $\Pro(Y_1 \leq t)$---where $t$ is the estimated quantile of of the reference group associated with specificity $p$. However, this effectively treats the reference distribution as if fixed and known in advance, and fails to account for duality in sampling variation. Instead, a point-wise confidence band can be formed by way of the nonparametric bootstrap. If $B$ denotes the number of bootstrap replicates (sufficiently large), the procedure is as follows in the case where $n_0$ and $n_1$ are fixed features of the study design:

\noindent \textbf{for} $b$ in 1 to $B$:
\begin{enumerate}
    \item Obtain a full-size resample (with replacement) of data from the reference group, $Y_{01}^{b}, \dots, Y_{0n_0}^{b}$. Let $\bF_{0b}^{-1}(p)$ denote the empirical quantile function of the bootstrapped observations.
    \item Obtain a full-size resample (with replacement) of data from the comparator group $Y_{11}^{b}, \dots, Y_{1n_1}^{b}$. Let $\bF_{1b}(t)$ denote the empirical CDF of the bootstrapped observations.
    \item Let $\roch_b(p) = \bF_{1b}(\bF_{0b}^{-1}(p))$ denote the value of the empirical ROC curve at the value of $p$.
\end{enumerate}
\textbf{end}\\

\noindent A 95\% confidence band for $\roc(p)$ can be based, for instance, on the 2.5$^{\text{th}}$ and 97.5$^{\text{th}}$ percentiles of $\lbrace \roch_b(p) \rbrace_{b = 1}^{B}$. When $n_0$ and $n_1$ are not fixed characteristics of the study design, a full (i.e., unconditional) bootstrap can be conducted in place of the above procedure (\citealp{Davison97}).

The estimated AUC associated with the empirical ROC curve can be computed using its connection to the Mann-Whitney \textit{U}-statistic (\citealp{Hanley82}). In particular, if $R$ denotes the sum of the ranks (i.e., of the pooled data of size $n$, and accounting for ties via averaging ranks) in either group. The estimated empirical AUC is given as follows:
\begin{eqnarray*}
    \widehat{\text{AUC}} = \frac{1 + |2(1 + (n_0 + 1)/(2n_1) - R/(n_0n_1)) - 1|}{2}.
\end{eqnarray*}
This formula is expressed to automatically follow the convention $0.5 \leq \text{AUC} \leq 1$. An R function is provided in Code Block 2 to compute the empirically estimated AUC.
\begin{lstlisting}[language=R]
## Code Block 2: Calculation of the Empirical AUC

## values : string of values for the measure of interest
## group  : string of indicators for group membership
compute_auc <- function(values, group) {
    n0 <- length(y0)
    n1 <- length(y1)
    R <- sum(rank(values, ties.method = "average")[group == 0])
    auc <- (1 + abs(2*(1 + (n0 + 1)/(2*n1) - R/(n0*n1)) - 1))/2
    return(auc)
}
## output : numeric value of AUC
\end{lstlisting}
A test of the weak null hypothesis, $H_0 : \text{AUC} = 0.5$, can be conducted using the Mann-Whitney test (e.g., \texttt{wilcox.test()} in R). The strong null hypothesis, $H_0 : \roc(p) = p$, can be tested nonparametrically using, for example, the Kolmogorov-Smirnov test (e.g., \texttt{ks.test()} in R; \citealp{Conover71}).

\subsection{Parametric estimation}

Suppose now that we are willing to posit a parametric form for the data, so that $Y_0$ has quantile function $F_0^{-1}(p; \btheta_0)$ and $Y_1$ has CDF $F_1(t; \btheta_1)$. If $\widehat{\btheta}_0$ and $\widehat{\btheta}_1$ denote, say, maximum likelihood estimates (MLEs) of $\btheta_0$ and $\btheta_1$ (respectively), then the ROC curve can be consistently estimated as $\roch(p) = F_1(F_0^{-1}(p; \widehat{\btheta}_0); \widehat{\btheta}_1)$ by the invariance property of MLEs.

\subsubsection{Exponential data}

Consider the parametric form of the biexponential ROC model described in Section 2.4, whereby $\widehat{\lambda}_0 = 1/\overline{Y}_0$ and $\widehat{\lambda}_1 = 1/\overline{Y}_1$ denote the MLEs of $\lambda_0$ and $\lambda_1$. To form a point-wise confidence interval for $\roc(p)$, we note the utility of the following variance-stabilizing transformation:
\begin{eqnarray*}
    \sqrt{n_0}(\log(\widehat{\lambda}_0) - \log(\lambda_0)) \overset{d}{\longrightarrow} \mathcal{N}(0, 1),
\end{eqnarray*}
so that, by independence of the observations, $\text{Var}(\log(\widehat{\lambda}_1) - \log(\widehat{\lambda}_0)) \approx n_0^{-1} + n_1^{-1}$ is an asymptotically valid approximation that lends itself to the following form for a point-wise confidence interval for $\roc(p)$ under the parametric biexponential model:
\begin{eqnarray*}
    1 - (1 - p)^{\exp\left(\left(\log(\widehat{\lambda}_1) - \log(\widehat{\lambda}_0)\right) \pm z_{1 - \alpha/2}\sqrt{n_0^{-1} + n_1^{-1}}\right)}.
\end{eqnarray*}
Notably, the point estimates and confidence intervals do not depend upon the specific values of $\widehat{\lambda}_1$ and $\widehat{\lambda}_0$ except through the value of $\widehat{\alpha} = \widehat{\lambda}_1/\widehat{\lambda}_0$, which is intuitive as the ROC curve is invariant to re-scaling the data. The biexponential ROC curve can reasonably be thought of as a one-parameter model. The AUC can be estimated as $\widehat{\text{AUC}} = 1 - (\widehat{\alpha} + 1)^{-1}$. Under the exponential model, a test of both the weak and the strong null can be implemented using the likelihood ratio statistic, $\Lambda = n\log(\widehat{\lambda}) - n_0\log(\widehat{\lambda}_0) - n_1\log(\widehat{\lambda}_1)$, where $\widehat{\lambda}$ denotes the MLE of the rate parameter in the combined data. The likelihood ratio test statistic has an asymptotic $\chi_1^2$ distribution under the strong null.

\subsubsection{Normal data}

Now, consider the parametric form of the binormal model described in Section 2.5, whereby $\widehat{\mu}_0 = \overline{Y}_0$, $\widehat{\mu}_1 = \overline{Y}_1$, $\widehat{\sigma}_0^2 = n_0^{-1}\sum_{i = 1}^{n_0} (Y_{0i} - \widehat{\mu}_0)^2$, and $\widehat{\sigma}_1^2 = n_1^{-1}\sum_{i = 1}^{n_1} (Y_{1i} - \widehat{\mu}_1)^2$; in practice, we may use the sample variances (e.g., $S_0^2 = n_0\widehat{\sigma}_0^2(n_0 - 1)^{-1}$) in place of the MLEs. Following a similar line of logic as with the exponential model reveals the following form for a point-wise confidence interval for $\roc(p)$ under the binormal model:
\begin{eqnarray*}
    \Phi\left( \frac{(\widehat{\mu}_0 - \widehat{\mu}_1) + S_0\Phi^{-1}(p) \pm t_{1 - \alpha/2, \text{df}}\sqrt{S_0^2n_0^{-1} + S_1^2n_1^{-1}}}{S_1}\right),
\end{eqnarray*}
where $\text{df}$ denotes the Welch-Satterthwaite approximation to degrees of freedom (\citealp{Welch47}). Similarly to the above example, the point estimates and confidence intervals do not depend upon the specific values of the estimates except through the values of $\widehat{\delta} = \widehat{\mu}_0 - \widehat{\mu}_1$ and $S_0/S_1$; the binormal ROC curve can be conceptualized as a two-parameter model. The AUC can be estimated numerically given the MLEs using, for example, the \texttt{integrate()} function in R. A test of the weak null can be based on an unequal-variance \textit{t}-test of means. On the other hand, a test of the strong null can be based on the likelihood ratio test, with $\Lambda = n\log(\widehat{\sigma}^2) - n_0\log(\widehat{\sigma}_0^2) - n_1\log(\widehat{\sigma}_1^2)$, where $\widehat{\sigma}^2$ denotes the MLE of the variance in the combined data. The likelihood ratio test statistic has an asymptotic $\chi_2^2$ distribution under the strong null (note, however, that the MLEs for $\sigma_0^2$ and $\sigma_1^2$ may not be replaced by the corresponding sample variances in the test statistic for this to hold).

\subsection{Semiparametric estimation}

Let $U_{ij} = I(Y_{1j} \leq Y_{0j})$ denote the indicator of the $j^{\text{th}}$ observation from the comparator group having a value no larger than the $i^{\text{th}}$ observation from the reference group (there are a total of $n_0 \times n_1$ observations of the form $U_{ij}$). It has been previously shown (\citealp{Pepe00}) that $\E[U_{ij}|F_0(Y_{0i}) = p] = \roc(p)$. This lends itself to the use of generalized linear models (suitable for binomial data) to estimate an ROC curve directly, bypassing the need to impose an explicit parametric structure.

\subsubsection{The semiparametric biexponential model}

Consider modeling the ROC curve using the biexponential form described in Section 2.4 (which is implied by, but does not itself imply, exponentially distributed data). Letting $V_{ij} = 1 - U_{ij}$, the corresponding semiparametric biexponential ROC model is given by $\log(\E[V_{ij}|F_0(Y_{0i}) = p]) = \alpha \log(1 - p)$. Note the absence of an intercept in this single-parameter family. In practice, the value of $p$ is not known but for each $V_{ij}$ is estimated as $\widehat{p}_{ij} = \bF_0(Y_{0i})$. The associated estimating equations for $\alpha$ take the following form:
\begin{eqnarray*}
    \mathbb{G}(\alpha) = \sum_{i = 1}^{n_0} \sum_{j = 1}^{n_1} \frac{\log(1 - \widehat{p}_{ij})}{1 - (1 - \widehat{p}_{ij})^\alpha}(V_{ij} - (1 - \widehat{p}_{ij})^\alpha) = 0.
\end{eqnarray*}
Observations with $\widehat{p}_{ij} = 1$ (so that $\log(1 - \widehat{p}_{ij}) = -\infty$) may be excluded for the estimating equation to be well defined; this will occur for the pairs that involve the maximum order statistic of the $Y_{0i}$'s, so that the estimating equation relies on $(n_0 - 1) \times n_1$ of the observations $U_{ij}$. An R function is provided in Code Block 3 to estimate and implement semiparametric biexponential estimation of the ROC curve.

\begin{lstlisting}[language=R]
## Code Block 3: Biexponential Semiparametric Estimation of the ROC curve

## values : string of values for the measure of interest
## group  : string of indicators for group membership
ROC.GLM.E <- function(group, values) {
     y0 <- values[group == 0]
     y1 <- values[group == 1]
     n0 <- length(y0)
     n1 <- length(y1)
     Pij <- matrix(colMeans(outer(y0, y0, "<=")), nrow = n0, ncol = n1)
     P <- matrix(Pij, ncol = 1)
     Q <- cbind(log(1 - P)[P != 1])
     V <- 1 - as.numeric(t(outer(y1, y0, "<=")))[P != 1]
     zz <- glm.fit(Q, V, family = binomial(link = "log"))
     zz$coefficients
}
## output : estimated coefficient from semiparametric biexponential ROC
\end{lstlisting}

This estimating equation possesses no closed-form solution, but can be solved numerically using a Gauss-Newton algorithm (\citealp{Mittelhammer00}). Due to the correlation of the $U_{ij}$'s, however, a bootstrap approach is typically employed in order to obtain asymptotically valid confidence intervals for $\roc(p)$. As in Section 3.2.1, the AUC can be estimated as $\widehat{\text{AUC}} = 1 - (\widehat{\alpha} + 1)^{-1}$. Both the weak and strong null can be tested using the bootstrap samples under an asymptotic normality assumption on $\widehat{\alpha}$.

\subsubsection{The semiparametric binormal model}

Now, consider modeling the ROC curve using the binormal form (which is implied by, but does not itself imply, normally distributed data). The corresponding semiparametric binormal ROC model is given by $\Phi^{-1}(\E[U_{ij}|F_0(Y_{0i}) = p]) = \beta_0 + \beta_1\Phi^{-1}(p)$. The associated estimating equations for $\bbeta = (\beta_0, \beta_1)^{\top}$ take the following form:
\begin{eqnarray*}
    \mathbb{G}(\bbeta) = \sum_{i = 1}^{n_0} \sum_{j = 1}^{n_1} \bX_{ij} \frac{\phi(z_{ij})}{\Phi(z_{ij})(1 - \Phi(z_{ij}))}(U_{ij} - \Phi(z_{ij})) = \textbf{0},
\end{eqnarray*}
where $z_{ij} = \beta_0 + \beta_1\Phi^{-1}(\widehat{p}_{ij})$, and $\bX_{ij} = (1, \Phi^{-1}(\widehat{p}_{ij}))^{\top}$.
For reasons analogous to those described in Section 3.3.1, observations with $\widehat{p}_{ij} = 1$ may be excluded for the estimating equations to be well defined. These estimating equations, too, possess no closed-form solutions, and a nonparametric bootstrap is typically employed to conduct inference. As in Section 3.2.2, the AUC can be estimated numerically given the MLEs using, for example, the \texttt{integrate()} function in R. The weak and/or strong null can be tested by conducting the appropriate tests on $\bbeta$, also previously described. An R function is provided in Code Block 4 to estimate implement semiparametric binormal estimation of the ROC curve.

\begin{lstlisting}[language=R]
## Code Block 4: Binormal Semiparametric Estimation of the ROC curve

## values : string of values for the measure of interest
## group  : string of indicators for group membership
ROC.GLM.N <- function(group, values) {
     y0 <- values[group == 0]
     y1 <- values[group == 1]
     n0 <- length(y0)
     n1 <- length(y1)
     Pij <- matrix(colMeans(outer(y0, y0, "<=")), nrow = n0, ncol = n1)
     P <- matrix(Pij, ncol = 1)
     Q <- cbind(1, qnorm(P)[P != 1])
     U <- as.numeric(t(outer(y1, y0, "<=")))[P != 1]
     zz <- glm.fit(Q, U, family = binomial(link = "probit"))
     zz$coefficients
}
## output : estimated coefficient from semiparametric binormal ROC
\end{lstlisting}

\subsection{Other considerations for characterizing uncertainty}

We acknowledged at the beginning of Section 3 that our outlined estimation procedures assume that it is of interest to characterize uncertainty in estimating the TPR at specific values of the FPR. It could be of equal interest to characterize uncertainty of the FPR at specific values of the TPR. We do not consider this nuance in extraordinary depth because estimation of this form can be accomplished by a convention change as described in Section 2.2 and illustrated in Figure 2, so that the fundamental mechanics of the procedure are unchanged. We acknowledge that the setting in which there is no clear ordering of groups, nor a natural way to define higher values of the measure of interest as ``better'' or ``worse'' poses a unique challenge. Others have considered intuitively combining information from bi-directional confidence regions, including radial sweeps, though there is little theoretical or empirical justification for these methods (\citealp{Horv08, Macskassy03}).

\section{Simulation-based comparison of estimation methods}

The goal of this section is to illustrate the bias-variance tradeoff associated with various approaches in finite samples. We utilize two general data generating mechanisms (DGMs): the first is an exponential DGM of the form $Y_0 \sim \text{Exponential}(\lambda_0 = 1)$ and $Y_1 \sim \text{Exponential}(\lambda_1 = 4)$, and the second is a normal DGM $Y_0 \sim \mathcal{N}(\mu_0 = 5.5, \sigma_0^2 = 1)$ and $Y_1 \sim \mathcal{N}(\mu_1 = 4.0, \sigma_1^2 = 1)$. We further consider two pairs of sample sizes: $n_0 = n_1 = 30$ (low) and $n_0 = n_1 = 60$ (high). In Section 4.1, we seek to confirm the inadequacy of the Wilson interval as compared to the nonparametric bootstrap for the empirical ROC curve. In Section 4.2, we generate data from the exponential DGM and compare the finite sample properties of the various methods described in Section 3. In Section 4.3, we do the same under the normal DGM.

To evaluate the finite sample properties of these methods in various scenarios, we compare the average estimated value of the ROC curve to the true value, and determine the average width and coverage associated with 95\% confidence intervals. For each simulation scenario presented, we utilize $M = 1000$ simulation replicates and, where applicable, $B=3000$ bootstrap replicates. All simulations were conducted in R (\citealp{R20}).

\vspace{-5mm}

\begin{table}[h!]
    \centering
    \caption{Results of a comparison between the Wilson interval and the quantile-based nonparametric bootstrap interval for the empirical ROC curve. $\roch(p)$ denotes the average estimate of the ROC curve at the specified value of $p$. $\Delta$ denotes the average confidence interval width and CP denotes coverage probability. ``-W'' denotes the Wilson interval and ``-B'' denotes the bootstrap interval.}
    \begin{tabular}{rcccccccccc}
        $n_0 = n_1$ & $p$ & $\roc(p)$ & $\roch(p)$ & $\Delta$-W & CP-W & $\Delta$-B & CP-B \\ \hline
        30 & 0.0027    & 0.1 & 0.31 & 0.29 & 0.31 & 0.51 & 0.30  \\ 
        30 & 0.0215    & 0.3 & 0.31 & 0.29 & 0.59 & 0.51 & 0.84   \\ 
        30 & 0.0670    & 0.5 & 0.55 & 0.32 & 0.73 & 0.59 & 0.96   \\ 
        30 & 0.1650    & 0.7 & 0.67 & 0.31 & 0.75 & 0.51 & 0.97   \\ 
        30 & 0.4140    & 0.9 & 0.90 & 0.22 & 0.94 & 0.27 & 0.96   \\  
        60 & 0.0027   & 0.1 & 0.22 & 0.19 & 0.37 & 0.38 & 0.40  \\ 
        60 & 0.0215   & 0.3 & 0.33 & 0.22 & 0.60 & 0.45 & 0.91  \\ 
        60 & 0.0670   & 0.5 & 0.52 & 0.24 & 0.72 & 0.45 & 0.96  \\ 
        60 & 0.1650   & 0.7 & 0.69 & 0.22 & 0.78 & 0.37 & 0.97  \\ 
        60 & 0.4140   & 0.9 & 0.89 & 0.15 & 0.89 & 0.20 & 0.98  \\    \hline
    \end{tabular}
    \label{tab:one}
\end{table}

\subsection{The Wilson interval vs. the bootstrap for the empirical ROC}

In this simulation study, we generate data from the normal DGM in order to compare the Wilson interval to the quantile-based nonparametric bootstrap-based interval associated with empirical ROC curve. We consider both the low and high sample size cases. The results of this simulation are shown in Table \ref{tab:one}, which depicts the results at the specified values of $p$ such that $\roc(p) = 0.1, 0.3, 0.5, 0.7$, and $0.9$; the justification for considering values of $p$ in this range that it encompasses a wide range of ordered pairs on the ROC curve that approach the boundary of the curve.

We observe a marked degree of finite-sample bias associated with the empirical ROC curve for very small values of $p$. The Wilson intervals uniformly have lower width as compared to those associated with the empirical bootstrap, which is not surprising given that it treats the distribution of the reference group as if fixed. With the exception of the lower values of $p$, the empirical bootstrap appears to possess adequate coverage. On the other hand, the Wilson interval generally does not possess proper coverage across the board. The intuition for the general pattern of under-coverage is that the Wilson interval fails to account for variation in estimating $p$, the specificity associated with a particular cut-off point. We no longer consider the Wilson interval as a competing method for the remainder of this manuscript.

\subsection{Comparisons under the exponential DGM}

In this section, we generated data under the exponential DGM, and considered four models: (1) a parametric binormal model, (2) a parametric exponential model, (3) a semiparametric biexponential model, and (4) a semiparametric binormal model. We consider both the low and high sample size cases. The results of this simulation are depicted in Figure 4. As expected, the parametric binormal ROC curve demonstrates a bias and a lack of coverage that are not overcome in a larger sample. The finite sample behavior is more favorable for the competing methods. The validity of the parametric and semiparametric biexponential models is not surprising (Figures 4(A) and 4(D)), nor is the apparent greater degree of efficiency associated with the parametric method (Figures 4(B) and 4(E)). It is noteworthy that the semiparametric binormal model exhibits so little bias given that it is not correctly specified, though we have pointed out that it offers greater flexibility as compared to the biexponential family. Given the additional model complexity of the binormal model, the finding that the confidence intervals are generally wider as compared to those of the respective biexponential models is not unexpected (Figures 4(B) and 4(E)).

\begin{figure}[h!]
    \centering
    \includegraphics[width=0.32\linewidth]{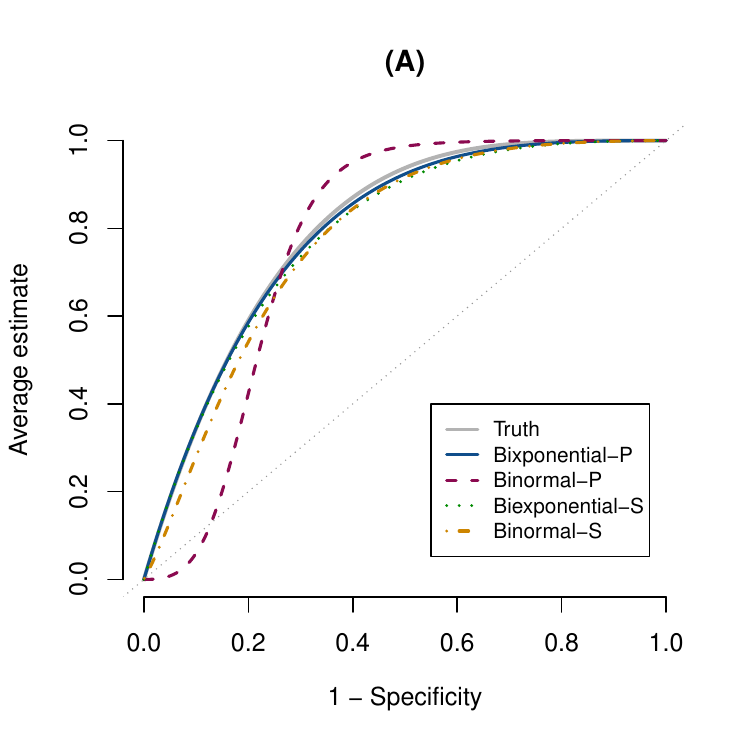}
    \includegraphics[width=0.32\linewidth]{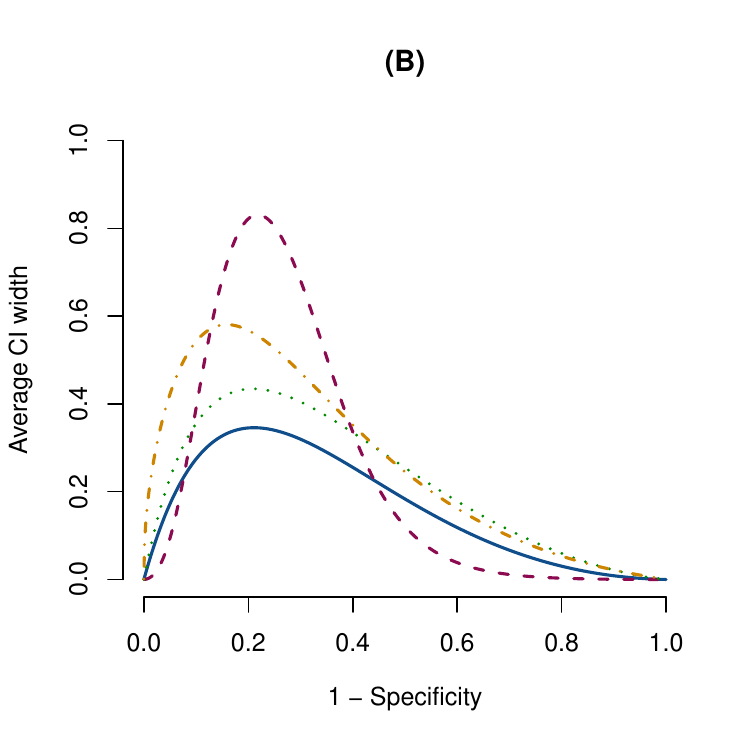}
    \includegraphics[width=0.32\linewidth]{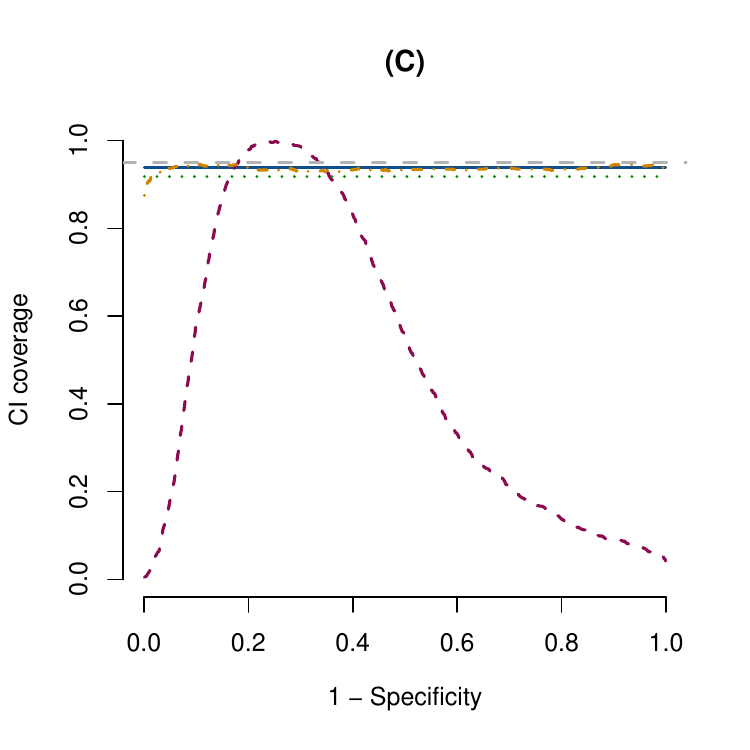}

    \includegraphics[width=0.32\linewidth]{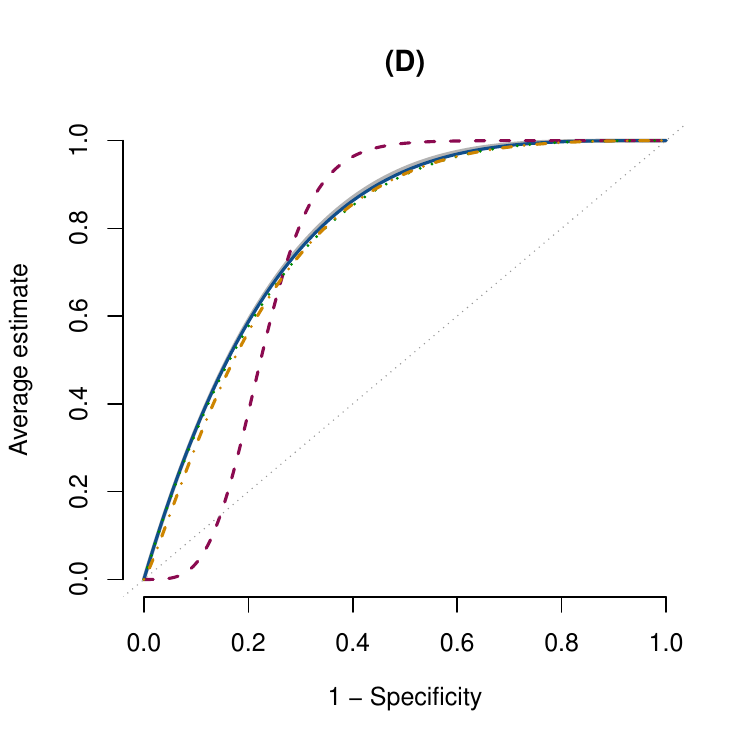}
    \includegraphics[width=0.32\linewidth]{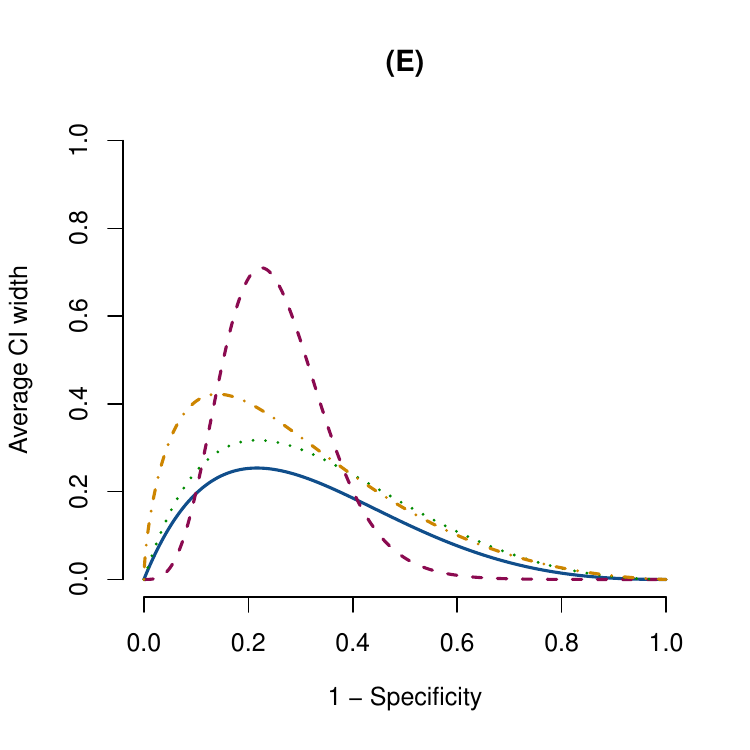}
    \includegraphics[width=0.32\linewidth]{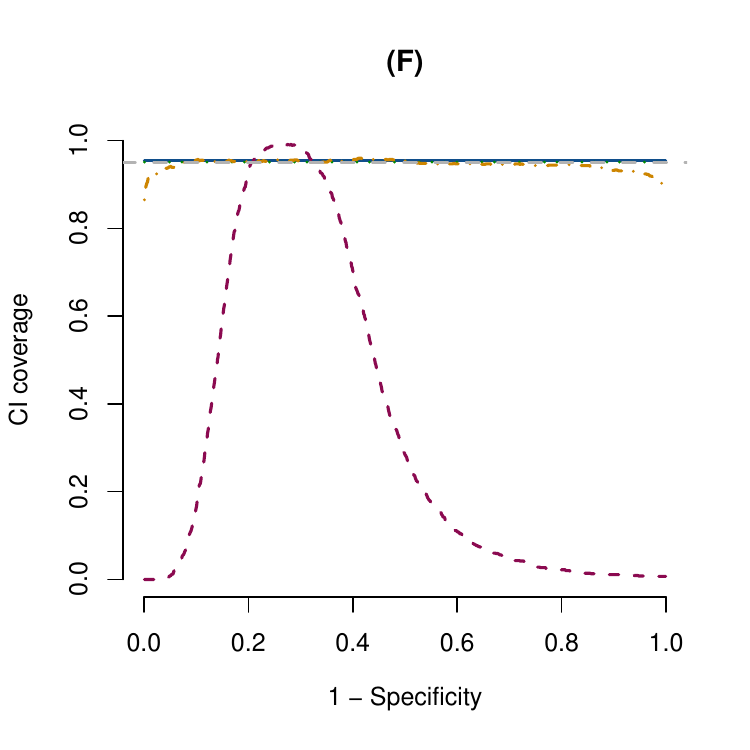}
    \caption{Results of a simulation study under the exponential DGM. Panels (A) through (C) correspond to the lower sample size ($n_0 = n_1 = 30$) and (D) through (F) correspond to the higher sample size ($n_0 = n_1 = 60$). Panels (A) and (D) depict the average estimated value of the ROC curve across methods, with the truth depicted as a reference. Panels (B) and (E) depict the average confidence interval width across $p$, the FPR. Panels (C) and (F) present coverage across methods, with a reference line of $y=0.95$.}
    \label{fig:Fig4}
\end{figure}

\begin{figure}[h!]
    \centering
    \includegraphics[width=0.32\linewidth]{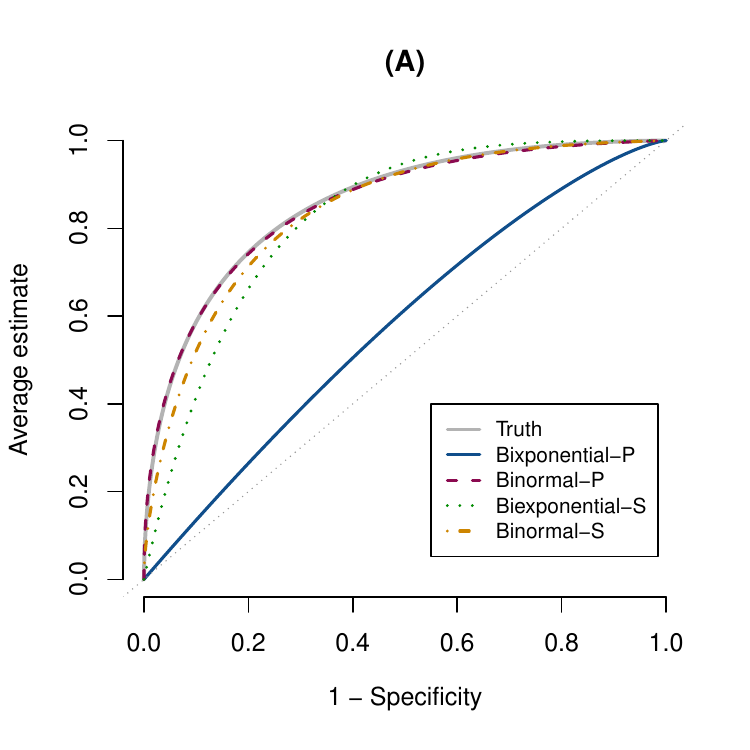}
    \includegraphics[width=0.32\linewidth]{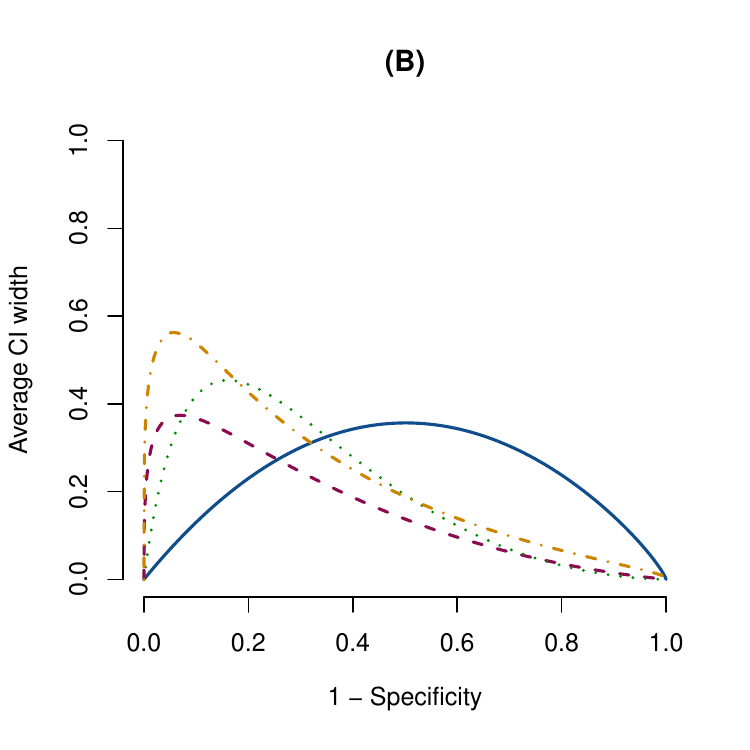}
    \includegraphics[width=0.32\linewidth]{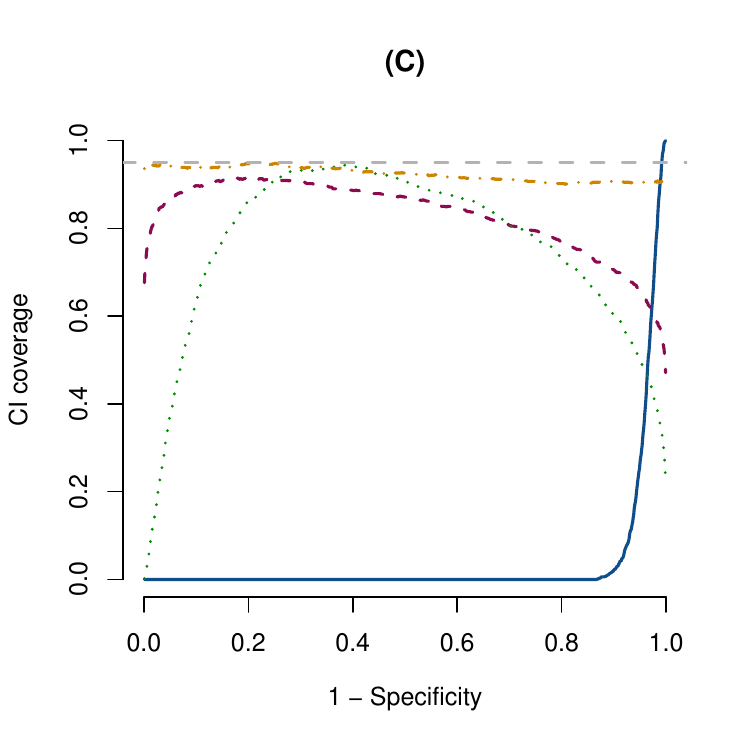}

    \includegraphics[width=0.32\linewidth]{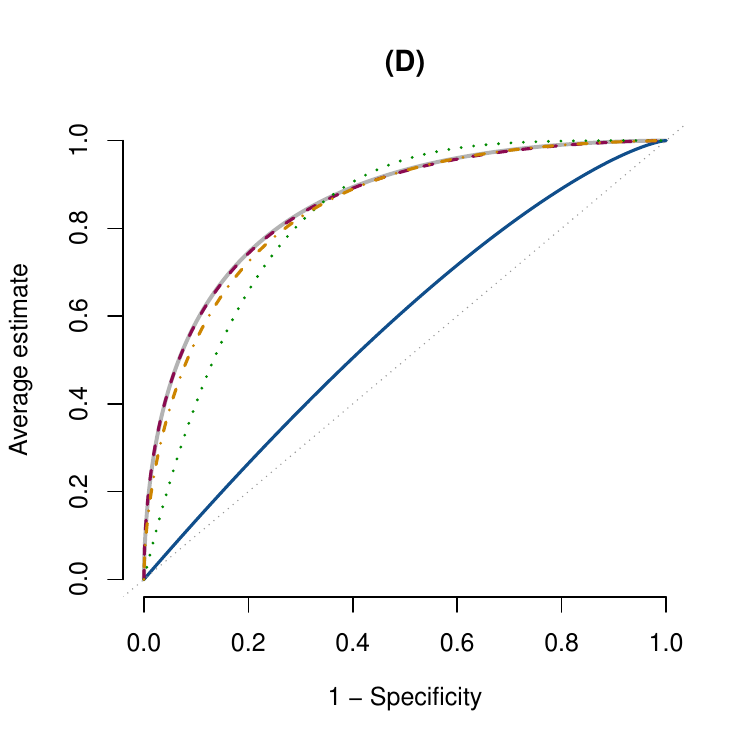}
    \includegraphics[width=0.32\linewidth]{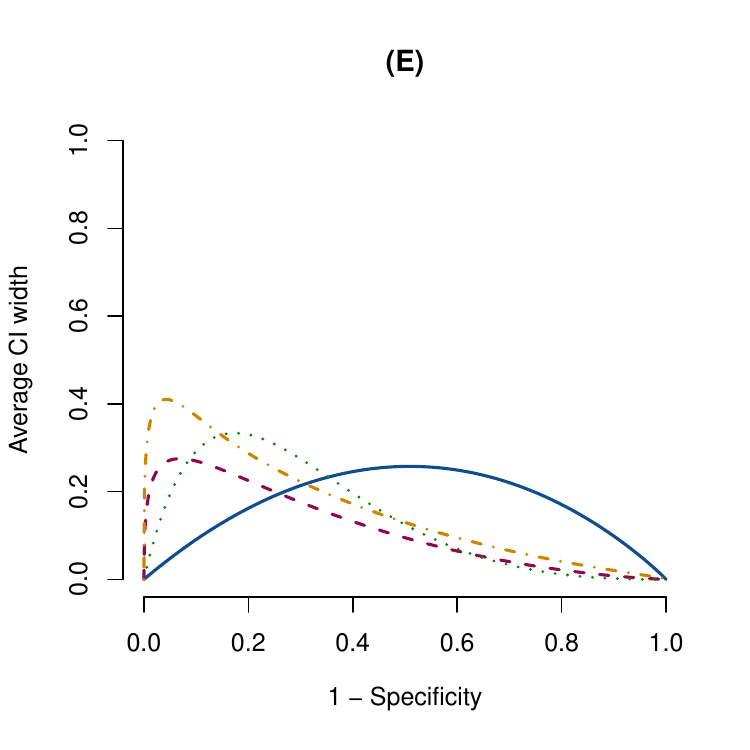}
    \includegraphics[width=0.32\linewidth]{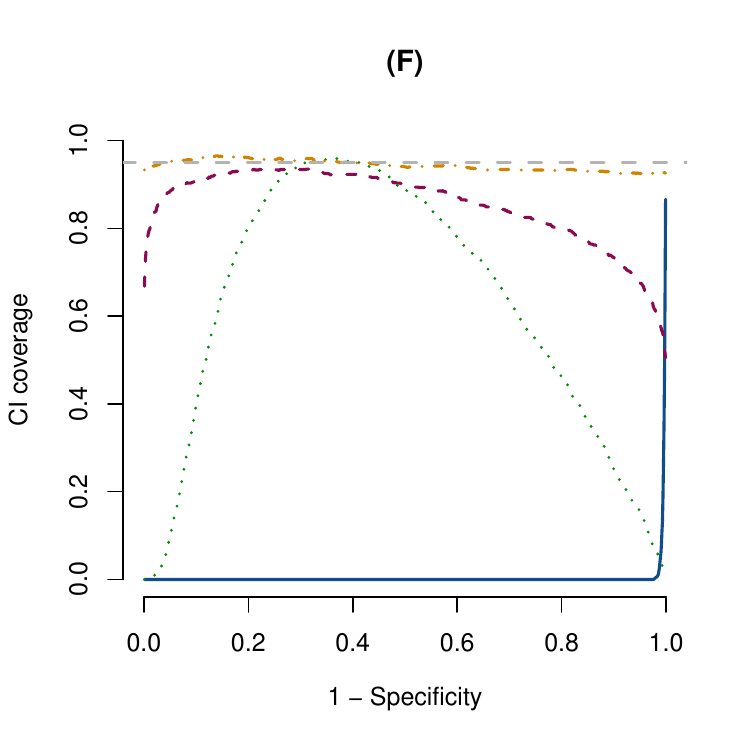}
    \caption{Results of a simulation study under the normal DGM. Panels (A) through (C) correspond to the lower sample size ($n_0 = n_1 = 30$) and (D) through (F) correspond to the higher sample size ($n_0 = n_1 = 60$). Panels (A) and (D) depict the average estimated value of the ROC curve across methods, with the truth depicted as a reference. Panels (B) and (E) depict the average confidence interval width across $p$, the FPR. Panels (C) and (F) present coverage across methods, with a reference line of $y=0.95$.}
    \label{fig:Fig5}
\end{figure}

\subsection{Comparisons under the normal DGM}

We conduct a simulation study that mirrors the setup described in Section 4.2, but instead generating data under the normal DGM. The results of this simulation are depicted in Figure 5. The parametric biexponential ROC curve performs poorly with respect to bias and coverage. This can be heuristically explained by the fact that the parametric biexponential model uses information from the mean ratio, despite the fact that the data are parameterized by the (scaled) mean difference and the ratio of standard deviations. Since the mean ratio of $\mu_1/\mu_0 = 5.5/4.0$ is close to one, the parametric biexponential model does not ``perceive,'' so to speak, a meaningful difference in stochastic ordering under the (albeit incorrect) exponential assumption. The finite sample behavior is more favorable for the competing methods, although the parametric binormal model suffers a notable loss of coverage near the boundaries. The semiparametric biexponential model is not free of bias, although its relatively good performance is likely related to the fact that the variances were generated to be equal in this data generating mechanism (that is to say that the biexponential form is strictly concave, whereas unequal variances in a normal model introduces infection points into the ROC curve). This study is consistent with the relative efficiency associated with the correct parametric specification of the data (Figures 5(B) and 5(C)).

\section{Application to solid organ transplants}

In this section, we apply several estimation methods to data from a study that sought to investigate SARS-CoV-2 vaccination-induced immunogenicity.

\subsection{Study cohort and measures}

Eligible individuals enrolled in this study included SOT recipients ($n_1 = 54$) and healthy controls (HCs; $n_0 = 26$) over 55 years old who were scheduled to receive a two-dose regimen of BNT162b2 as part of routine care at Vanderbilt University Medical Center (Nashville, TN). Specific transplantation groups included prior heart (11\%), kidney (41\%), liver (37\%), lung (7\%), and kidney/liver (4\%) transplant recipients. Immunoglobulin G to SARS-CoV-2 spike extracellular domain (ECD) was evaluated by enzyme-linked immunosorbent assay. Three SOT recients were excluded from our analysis due to serological evidence of an ongoing or prior SARS-CoV-2 infection. Further details pertaining to recruitment, enrollment, and prior analysis are detailed by \citet{Yanis22}.

\subsection{Empirical analysis}

Our illustrative analysis specifically focuses on ECD response to the second vaccination dose (henceforth denoted $\ecd$). Higher values of $\ecd$ serve as a correlate of higher immunity (\citealp{Oppenheimer23}), and we anticipate \textit{a priori} that SOT recipients, due to immunocompromised status, tend to have lower $\ecd$ values as compared to HCs. Figure 6(A) presents histograms of $\ecd$, stratified by group, and Figure 6(B) presents an empirical ROC curve with a 95\% confidence band as estimated by the nonparametric bootstrap procedure described in Section 3.1. The degree of separation between the distribution of groups is apparent from examining the histograms. The empirical ROC curve, particularly when considering the associated confidence bands, offers additional context regarding our certainty regarding how well-separated the groups are in their $\ecd$ response.

\begin{figure}[h!]
    \centering
    \includegraphics[width=0.77\linewidth]{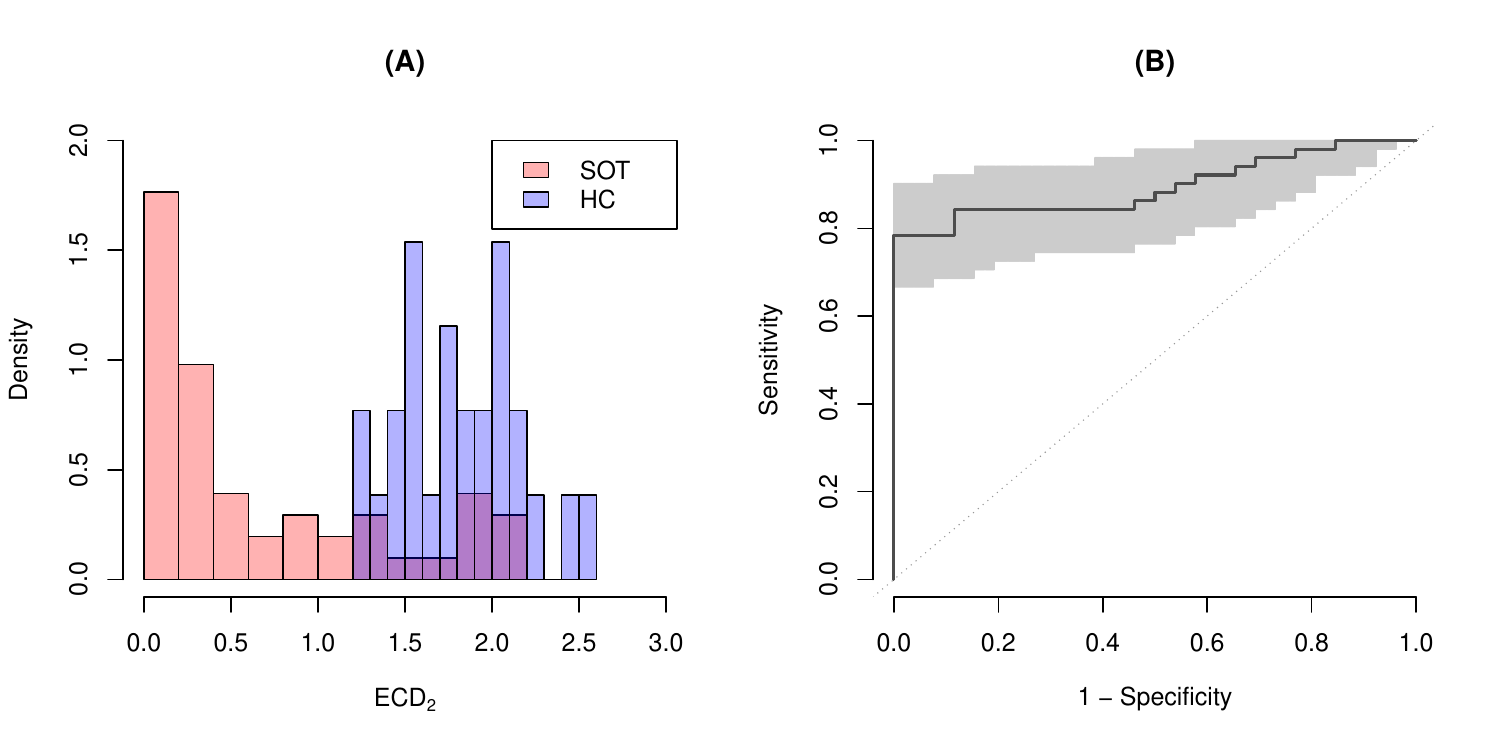} \\
    \includegraphics[width=0.77\linewidth]{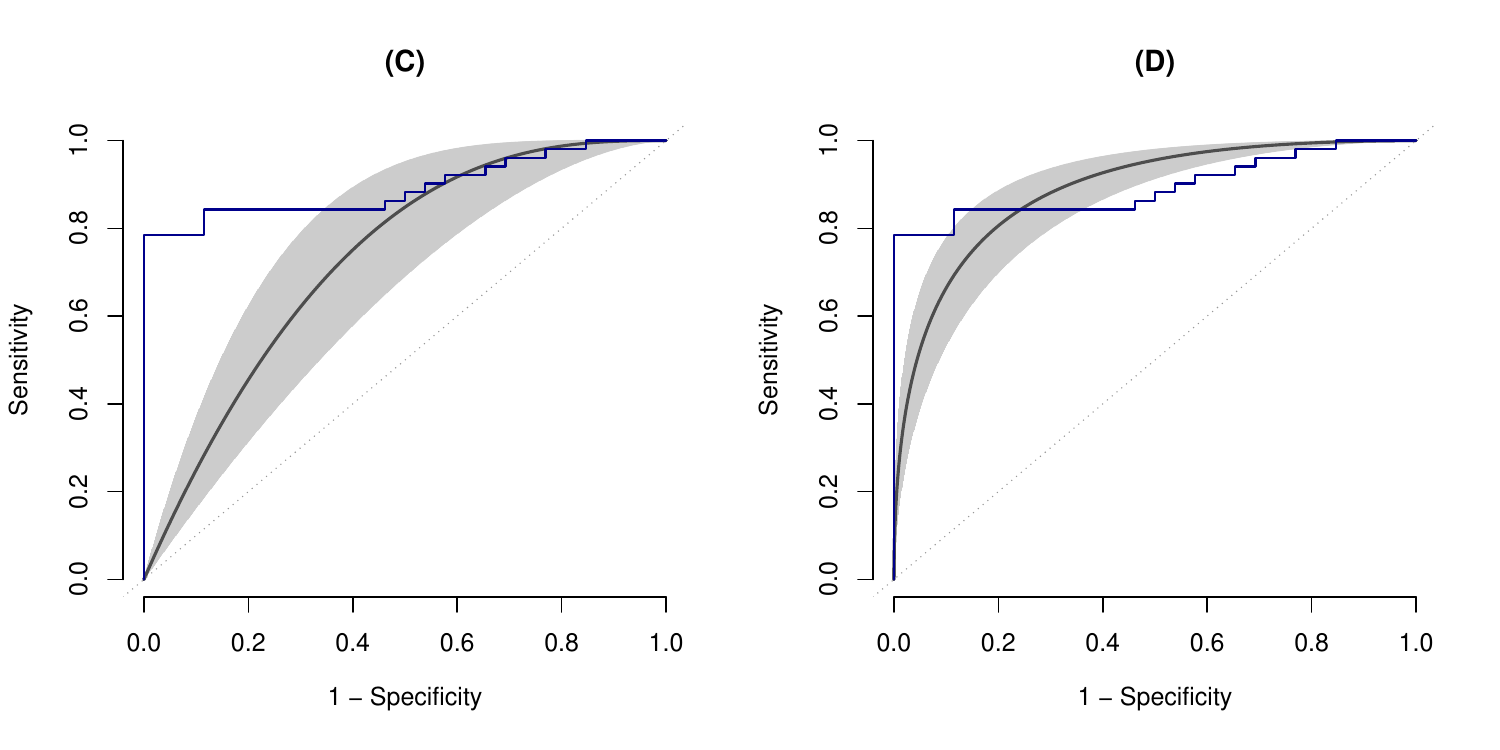}\\
    \includegraphics[width=0.77\linewidth]{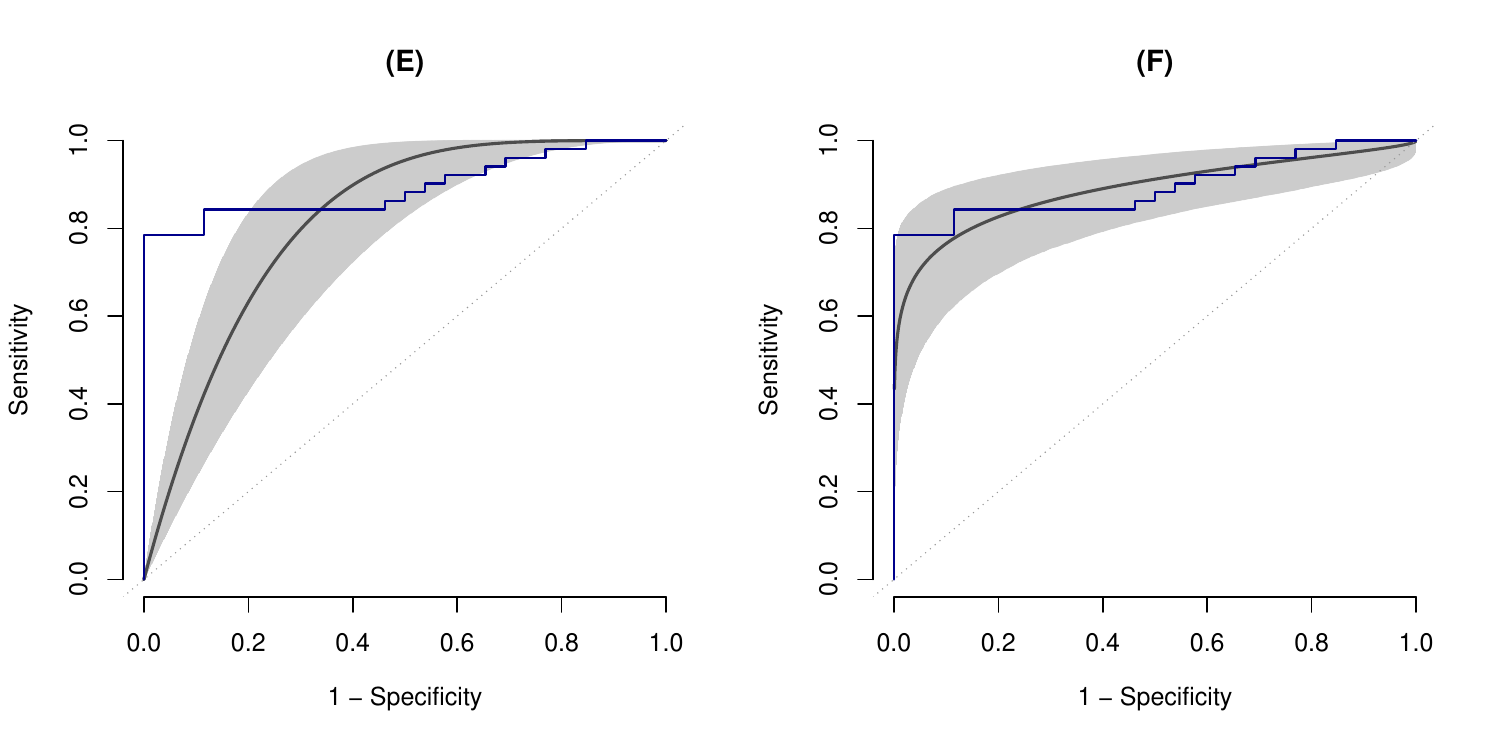}
    \caption{Panel (A) depicts histograms of $\ecd$. Panel (B) depicts the empirical ROC curve with a confidence band. Panels (C) through (F) depict various ROC estimates, including point estimates (dark gray solid line) and confidence bands (light gray); the empirical ROC curve is also depicted in solid blue as a reference. Panel (C): parametric biexponential. Panel (D): parametric binormal. Panel (E): semiparametric biexponential method. Panel (F) semiparametric binormal.}
    \label{fig:Fig6}
\end{figure}

\clearpage

\subsection{Parametric estimation}

The estimated ROC curve under the parametric biexponential model is depicted in Figure 6(C), and the estimated curve under the parametric binormal model is depicted in Figure 6(D). For the parametric biexponential model, we estimate $\widehat{\alpha} = 2.72$. For the binormal model, we estimate $\widehat{\beta}_0 = 1.70$ and $\widehat{\beta}_1 = 0.52$. Of note, the distribution of $\ecd$ among SOT recipients exhibits right-skewness, whereas the distribution among HCs more closely resembles the symmetry of a normal distribution. That is to say that results from both parametric methods are likely biased, which is reflected in part by the discrepancy between the point estimates for the estimated ROC curves and the empirical curve; this discrepancy is particularly notable for the parametric biexponential model.

\subsection{Semiparametric estimation}

We now estimate the ROC curve using the correspoinding semiparametric methods. The results of the semiparametric biexponential model are shown in Figure 6(E) and those of the semiparametric binormal model are shown in Figure 6(F). For the biexponential model, we estimate $\widehat{\alpha} = 4.48$, a marked difference from the point estimate obtained using parametric methods. Recall that $\alpha$ does not possess the interpretation of a mean ratio when the biexponential curve is estimated in a semiparametric fashion. We see from comparing Figure 6(E) to 6(C) that the semiparametric biexponential model---while still notably different from the empirical curve---is better calibrated to the empirical curve across the broader range of $p$.

For the semiparametric binormal model, we estimate $\widehat{\beta}_0 = 1.35$ and $\widehat{\beta}_1 = 0.49$. A particularly noteworthy result from the semiparametric binormal fit is that the empirical ROC curve is almost entirely captured within its confidence band; this is not true for any of the other three fits. The curve associated with the parametric model possesses a similar shape, but likely overstates certainty. This example is again consistent with the flexibility associated with the semiparametric binormal curve, especially as compared to the more parsimonious biexponential model.

One key comparison is also between the semiparametric biexponential model (Figure 6(E)) and the parametric binormal model (Figure 6(D)). The confidence bands are considerably wider under the former model, despite the fact that it has one fewer parameter. This might be a more surprising result if had the two models been nested, but this particular comparison is a stark example of the fact relative efficiency is driven by more than simply the relative number of parameters possessed by the model. One possible source of this discrepancy is a ``ceiling effect'' in the variability given the boundedness of the ROC function.

\section{Discussion}

In this manuscript, we have described the ROC curve and highlighted aspects of its geometry that we believe may be underappreciated. We have outlined and offered software code for a number of competing nonparametric, parametric, and semiparametric estimation approaches. Simulation studies revealed, as expected, a notable trade-off between efficiency and bias associated with structure.

We acknowledge that the ROC curve may, for many, serve as a visual tool to complement a primary analysis and that explicitly conducting inference on the value of an ROC curve at a specified value of $p$ may not always reflect the main scientific goal. However, the width of confidence bands offer visual clarity to uncertainty and speak in some respects to how trustworthiness of the findings regarding diagnostic utility or stochastic ordering. Our simulations and application revealed non-trivial differences in both bias and efficiency across approaches. To that end, knowledge regarding the bias and efficiency of the approach used to generate an ROC curve is relevant even if confidence intervals for the curve are not specifically generated in a given researcher's implementation.

We note that loss of coverage is a commonly encountered feature near the boundary of the ROC curve. This may not always be of major concern in the sense that the extreme ends of the curve signify either the setting of zero-specificity or zero-sensitivity, reflecting cut-off points that may not be considered viable candidates. However, further investigation may bring clarity to the downstream consequences on coverage associated with confidence intervals for other summary measures (say, the AUC).

Our finding that the binormal estimation techniques---parametric and semiparametric---appear to offer a certain degree of robustness, is consistent with previous findings (\citealp{Hanley88, Devlin13, Bandos17}). Heuristically, the space of monotone increasing functions from $[0, 1]$ to $[0, 1]$ likely does not need to be indexed by more than two parameters in order to cover a wide range of real-world scenarios. Our work contributes to the literature by underscoring the inadequacy of the single-parameter biexponential model. The semiparametric variant of the biexponential model has previously been compared to other methods (\citealp{Devlin13}), though in the context of estimating $\alpha$ rather than $\roc(p)$ at various values of $p$. Note that the findings of this work do not necessarily support the uptake of the semiparametric biexponential model in the real world.

In our application of ROC methods to vaccine-associated immunogenicity in Section 5, there is a notable degree of heterogeneity in the comparator group that is not captured by our illustrative analysis. For example, liver transplant recipients undergo the lowest degree of immunosuppression as compared to other organ transplant types (\citealp{Pilch21}). Over the years, there has been research on regression of the ROC curve, whereby one seeks to evaluate how the nature of an ROC curve varies across population strata (\citealp{Pepe98, Alonzo02}). The techniques used to accomplish this are a natural extension of those presented in Section 3.3. This area has also been explored independent of the setting of diagnostic markers (\citealp{Illenberger22}).

We acknowledge that the suite of methods compared in this manuscript are not exhaustive. For example, \citet{Wan07}, \citet{Kim13}, and \citet{Ghebremichael24} consider alternative semiparametric methods for ROC estimation based on accelerated regression models and/or kernel density estimation. The goal of this work was not to serve as an exhaustive literature review of all existing ROC estimation techniques, but instead to provide insights into ROC analysis based on a selection of estimation approaches that are straightforward to implement in the real world. We hypothesize that the patterns of behavior for other semiparametric approaches would in many ways mirror those of the approaches we considered, although further study would be needed to confirm this.

An additional trade-off that we did not consider in this work is that of computational efficiency. The bootstrap is a computationally taxing procedure, particularly when one needs to employ enough replicates to justify quantile-based confidence intervals. Alonzo and Pepe consider coarsening the grid of FPR values at which to evaluate the TPR; more variability is introduced when failing to leverage the full distribution of the reference group (\citealp{Alonzo02}). In the modern era of improving computing resources, this trade-off may seem to wane in relevance; however, we are also in a modern era of access to increasing larger data sets, so considerations regarding the trade-off of computational burden should not be totally ignored.

We must offer our full acknowledgement that ROC methods are not without limitations. Some of the most meaningful criticisms of the methodology include that the ROC curve gives consideration (both visually and numerically via the AUC) to the non-viable region $[0.5, 1] \times [0, 0.5]$ in which both the specificity and sensitivity are seriously lacking, and that the ROC curve offers no meaningful insights into the positive predictive value and negative predictive value (\citealp{Lobo08}). These challenges have not gone completely unaddressed in the literature. For example, weighted AUC methods have been proposed to more heavily consider viable regions (\citealp{Li09}).

There are a number of important future directions for this work. Considerations regarding study design, matching, and repeated measures remain under-explored in this space. However, if the trends of interest in measures of stochastic ordering continue, we will soon identify unmet needs for continued expansions and extensions of ROC methodology to accommodate the nature of real-world data collection.

\acks{The Vanderbilt Institute for Clinical and Translational Research (VICTR) is funded by the National Center for Advancing Translational Sciences (NCATS) Clinical Translational Science Award (CTSA) Program, Award Number 5UL1TR002243-03. The content is solely the responsibility of the authors and does not necessarily represent the official views of the NIH.}

\bibliography{bib}

\end{document}